\documentclass{webofc}

\usepackage{color}
\usepackage{amsmath}
\usepackage{mathtools}  
\usepackage[varg]{txfonts}   
\usepackage{xcolor}
\usepackage{caption}
\usepackage{subcaption}  
\usepackage{hyperref}
\usepackage{csquotes} 

\newcommand{\pt}{p_\text{T}}    
\newcommand{\effDmPt}[1]{\epsilon^\text{DM}_{\pt>#1}}
\newcommand{\effPerfectPt}[1]{\epsilon^\text{perfect}_{\pt>#1}}
\newcommand{\effLhcPt}[1]{\epsilon^\text{LHC}_{\pt>#1}}
\newcommand{\ntracks}{N_{\text{t}}}
\newcommand{\relu}{\operatorname{ReLU}}

\newcommand{\ecless}{\leq_\text{EC}}

\newcommand{\gev}{\mathrm{GeV}}
\begin{document}
\title{An Object Condensation Pipeline for Charged Particle Tracking at the High Luminosity LHC}
%
%

\author{
    \firstname{Kilian} \lastname{Lieret}\inst{1,2}\fnsep\thanks{\email{kl5685@princeton.edu}} 
    \and
    \firstname{Gage} \lastname{DeZoort}\inst{1}\fnsep\thanks{\email{jdezoort@princeton.edu}}
}

\institute{Princeton University\and Institute for Research and Innovation in Software for High Energy Physics (IRIS-HEP)}

\abstract{%
Recent work has demonstrated that graph neural networks (GNNs) trained for charged particle tracking can match the performance of traditional algorithms while improving scalability to prepare for the High Luminosity LHC experiment. Most approaches are based on the edge classification (EC) paradigm, wherein tracker hits are connected by edges, and a GNN is trained to prune edges, resulting in a collection of connected components representing tracks. These connected components are usually collected by a clustering algorithm and the resulting hit clusters are passed to downstream modules that may assess track quality or fit track parameters.
In this work, we consider an alternative approach based on object condensation (OC), a multi-objective learning framework designed to cluster points belonging to an arbitrary number of objects, in this context tracks, and regress the properties of each object. 
We demonstrate that OC shows very promising results when applied to the pixel detector of the trackML dataset and can, in some cases, recover tracks that are not reconstructable when relying on the output of an EC alone.
The results have been obtained with a modular and extensible open-source implementation that allows us to efficiently train and evaluate the performance of various OC architectures and related approaches.}
\maketitle
\section{Introduction}
\label{intro}

The exploration of tracking algorithms based on graph neural networks (GNNs) is motivated by the poor computational scaling of combinatorial Kalman filter algorithms with pileup~\cite{Cerati_2016}. 
In recent years, many approaches have been developed around the edge classification (EC) paradigm (see e.g., Ref.~\cite{ju_exatrkx_2021}), in which GNNs are designed to predict whether or not edges drawn between tracker hits represent physical trajectories. These architectures have been shown to demonstrate excellent physics performance and, importantly, scalability with respect to pileup~\cite{lazar_accelerating_2023}.
In the majority of these approaches, tracks are rendered from edge-weighted graphs directly, either by a graph walk algorithm, spatial clustering, or simply collecting connected components. 

This work instead explores a learned track rendering stage based on object condensation (OC), a multi-loss training scheme we use to cluster hits belonging to the same track in a learned clustering space.
Employing only a very lightweight EC network (without any message passing), we show that the OC approach is able to deliver excellent performance when applied to simulated high-pileup TrackML pixel detector events.
We also show that the algorithm is able to reconstruct tracks with missing edges, which is not possible when relying on the output of an EC alone.
Therefore, this algorithm could also be used at the end of an EC pipeline, leveraging both node and edge embeddings to resolve ambiguities that may exist at the output of the EC.
\section{Dataset and input features}
\label{sec:dataset}
This study is performed using the TrackML dataset~\cite{amrouche_tracking_2020,amrouche_tracking_2021} that simulates the worst-case HL-LHC pileup conditions ($\langle\mu\rangle=200$) in a generic tracking detector geometry\footnote{We use the version corresponding to throughput phase of the TrackML challenge, which is hosted at Codalab~\cite{amrouche_tracking_2021}.}.
Our studies are limited to the innermost pixel detector layers, including 4 barrel layers and 7 layers in each endcap. 
In each pixel tracker event, we embed hits as graph nodes with 14 features, including:
\begin{itemize}
    \item 
    The cylindrical coordinates $r$, $\phi$, $z$ of the hits, where $z$ corresponds to the line of the colliding proton-proton pairs; the corresponding pseudorapidity $\eta$ and the conformal tracking coordinates $u$ and $v$~\cite{Hansroul_conformal_1988} are also included. 
    \item 
    The hit's charge fraction (sum of charge in the cluster divided by the number of activated channels), as well as variables describing the shape and orientation of the cluster introduced in Ref.~\cite{fox4DTrackingUsing2021}. For the latter, we closely follow the implementation in Ref.~\cite{ju_exatrkx_2021}.
\end{itemize}
\section{Tracking metrics}
\label{sec:metrics}
We study several tracking definitions to evaluate the performance of our pipeline. They are defined with respect to target populations of tracks; for example, we commonly report the tracking efficiency on \textit{reconstructable particles} that produce at least three hits and have $|\eta|<4.0$. 
The efficiencies are defined with respect to various matching criteria between reconstructed tracks and truth tracks.
For a given reconstructed track $c$, we define the \emph{majority particle} $\pi_c$ as the particle with the largest number of hits within $c$ (and choose a random one if this condition applies to multiple particles). 
We write $\#c$ for the number of hits in $c$ and $\#\pi_c$ for the number of hits of $\pi_c$ anywhere.
We define the \emph{majority fraction} $f^\text{in}_c$ as the number of hits of $\pi_c$ within $c$ divided by $\#c$ and the \emph{majority outside fraction} $f^\text{out}_c$ as the number of hits of $\pi_c$ outside of $c$ divided by $\#\pi_c$.
\begin{itemize}
    \item \textbf{Perfect match efficiency} ($\epsilon^\text{perfect}$): the number of reconstructed tracks with $\#c>3$, $\pi_c$ reconstructable, $f^\text{in}_c=f^\text{out}_c=1$ normalized to the number of reconstructable particles.
    \item \textbf{LHC-style match efficiency} ($\epsilon^\text{LHC}$): the number of reconstructed tracks with $\#c>3$, $\pi_c$ reconstructable, $f_c>75\%$ normalized to the number of clusters with reconstructable $\pi_c$. Note that duplicates, wherein multiple reconstructed tracks match to one particle, are possible with this definition. 
    \item \textbf{Double majority match efficiency} ($\epsilon^\text{DM}$): the number of reconstructed tracks with $\#c>3$, $\pi_c$ reconstructable, $f^\text{in}_c>50\%$, and $f^\text{out}_c>50\%$ normalized to the number of reconstructable particles. This definition produces unique cluster-track assignments. 
\end{itemize}
We also define the fake rate $f^\text{DM}$ based on the double majority metric as the number of reconstructed tracks with $\#c\geq 3$ and $\pi_c$ reconstructable that do not satisfy the double majority criterion normalized to the number of clusters with reconstructable $\pi_c$.

As we are mostly interested in high-$\pt$ tracks, we also consider these metrics with an additional $\pt>0.9\,\gev$ threshold applied to particles and majority particles in the definition of the metrics. 
The corresponding metrics are denoted $\effDmPt{0.9}$, $\effPerfectPt{0.9}$, $\effLhcPt{0.9}$ and $f^\text{DM}_{\pt>0.9}$.
\section{Graph construction}
The edges of the input graph are constructed by connecting hits on different detector layers that satisfy a series of geometric constraints and pass a classifier threshold.
\subsection{Edges based on geometric constraints}
\label{sec:geoconstraints}
This procedure is nearly identical to the geometric graph construction procedure described in Ref.~\cite{dezoort_interaction_2021}, but without applying any cuts based on truth information. 
Edges between nodes $i$ and $j$ are selected based on the following geometric variables:
\begin{gather}
    z_0 \coloneqq z_i - r_i \frac{z_j-z_i}{r_j-r_i},  \quad
    \phi_\text{slope} \coloneqq \frac{\phi_j-\phi_i}{r_j-r_i} , \quad\text{and}\quad
    \Delta R  \coloneqq  \sqrt{(\eta_j-\eta_i)^2 + (\phi_j-\phi_i)^2}.
\end{gather}
Here, we require candidate edges to satisfy $z_0 < 197.4\,\mathrm{mm}$, $\phi_\text{slope}< 0.001825/\mathrm{mm}$, and $\Delta R< 1.797$.
The cutoff points were optimized to maximize the fraction of reconstructable track edges appearing in the graph, while simultaneously minimizing the number of un-physical edges constructed. 
In contrast to Ref.~\cite{dezoort_interaction_2021}, no barrel intersection cut is applied.
Note that the performance of the graph construction can be translated to an approximate upper bound for the performance of the pipeline downstream.
Assuming that the pipeline can build tracks exactly out of those hits that are connected, that is, assuming a pipeline with perfect EC followed by identifying tracks as connected components of the resulting edge subgraph, we obtain $\effDmPt{0.9} \ecless 97.4\%$ and $\effPerfectPt{0.9} \ecless 84.0\%$.
However, we will also show that OC can, in some cases, surpass this limit.
%

%
We provide four initial edge features based on the coordinates of the two hits involved: $\Delta r$, $\Delta \phi$, $\Delta z$, and $\Delta R$.
The resulting graphs are denoted $\mathcal G\coloneqq (X, R_a, I)$, where $X\coloneqq (x_i)_{i=1,\dots,N}\in\mathbb{R}^{N\times 14}$ are the node features, $I\in\mathbb N^{2\times N_\text{edges}}$ is the list of edges in coordinate format, and $R_a\coloneqq (e_{ij})_{(i,j)\in I}$ with $e_{ij}\in \mathbb R^4$ are the edge features\footnote{While edge attributes and features are vectors of length $N_\text{edges}$, they are indexed by double indices $(i, j)\in I$ (abbreviated to $ij$ in subscript) for notational convenience.}.
We also define truth labels $l_i\in\{0,1,...,\ntracks\}$ (where $\ntracks$ is the number of particles in the graph) indicating whether the hit is noise ($l_i=0$) or belongs to track $t$ ($l_i=t$, $1\leq t\leq \ntracks$); in this work, we do not consider shared hits between tracks. 
The truth label $y_{ij}$ indicates  whether an edge connects two non-noise hits of the same particle ($l_i=l_j>0$, $(i, j)\in I$).
%
%
The geometric cuts alone achieve a purity of $N_\text{true}^\text{built} / N_\text{total}^\text{built}=4.5\%$ at $2.8\times 10^6$ edges per graph.
\subsection{Edge filtering}
\label{sec:edge-filtering}
We then apply a lightweight EC to reduce the number of false edges. 
For this, we train a fully connected neural network (FCNN) that takes node and edge features as inputs. 
The node and edge features described in~\autoref{sec:dataset} and~\autoref{sec:geoconstraints} are concatenated, $z^{(0)}_{ij}\coloneqq [x_i, x_j, e_{ij}]$, $(i,j)\in I$, and embedded into a 256-dimensional space by a fully connected layer: $z^{(1)}_{ij} \coloneqq W^{(1)} z^{(0)}_{ij}$, with learnable weights $W^{(1)}\in\mathbb{R}^{256\times(14+14+4)}$. 
We then apply a fully connected network of five hidden layers of width 256 with $\relu$ activations and residual connections of the form $z^{(\ell+1)}_{ij} \coloneqq \sqrt{\beta}\, W^{(\ell+1)} \relu\bigl(z^{(\ell)}_{ij}\bigr) +  \sqrt{1-\beta}\, z^{(\ell)}_{ij}$, where $l=1,\dots,5$, $(i, j)\in I$, and  $\beta=0.4$.

To obtain an edge weight, we apply the logistic activation function $\sigma$: $w_{ij}\coloneqq \sigma(W^{(7)}\relu(z^{(6)}_{ij}))\in(0,1)^\mathrm{N_\mathrm{edges}}$, where $W^{(7)}\in\mathbb R^{1\times 256}$.
This output is trained with binary cross entropy loss to classify whether an edge connects two hits of the same particle. 
As we are more interested in tracks with a high value of $\pt$, we exclude true edges connecting hits of particles with $\pt < 0.9\,\gev$ from the loss, i.e.,
\begin{align}
\begin{split}
    &\ell_{\text{EF}}(y, w) \coloneqq 
        -\frac 1{N_\text{edges}}\sum_{(i,j)\in I} \biggl( \delta_{(\pt > 0.9)}\, y_{ij} \log w_{ij} + (1-y_{ij})\log (1-w_{ij})\biggr),\\
    &\qquad\text{where}\quad \delta_{(\pt > 0.9)} \coloneqq \begin{cases}
        0 & l_i =0 \lor p_{T}^{l_i} < 0.9\,\gev,\\
        1 & \text{else}
    \end{cases},
\end{split}
\label{eq:ef_loss}
\end{align}
and $\pt^{l_i}$ denotes the $\pt$ of the particle belonging to particle~$i$.

\begin{figure}
    \centering
    \begin{subfigure}{0.52\linewidth}
    \includegraphics[height=4.8cm]{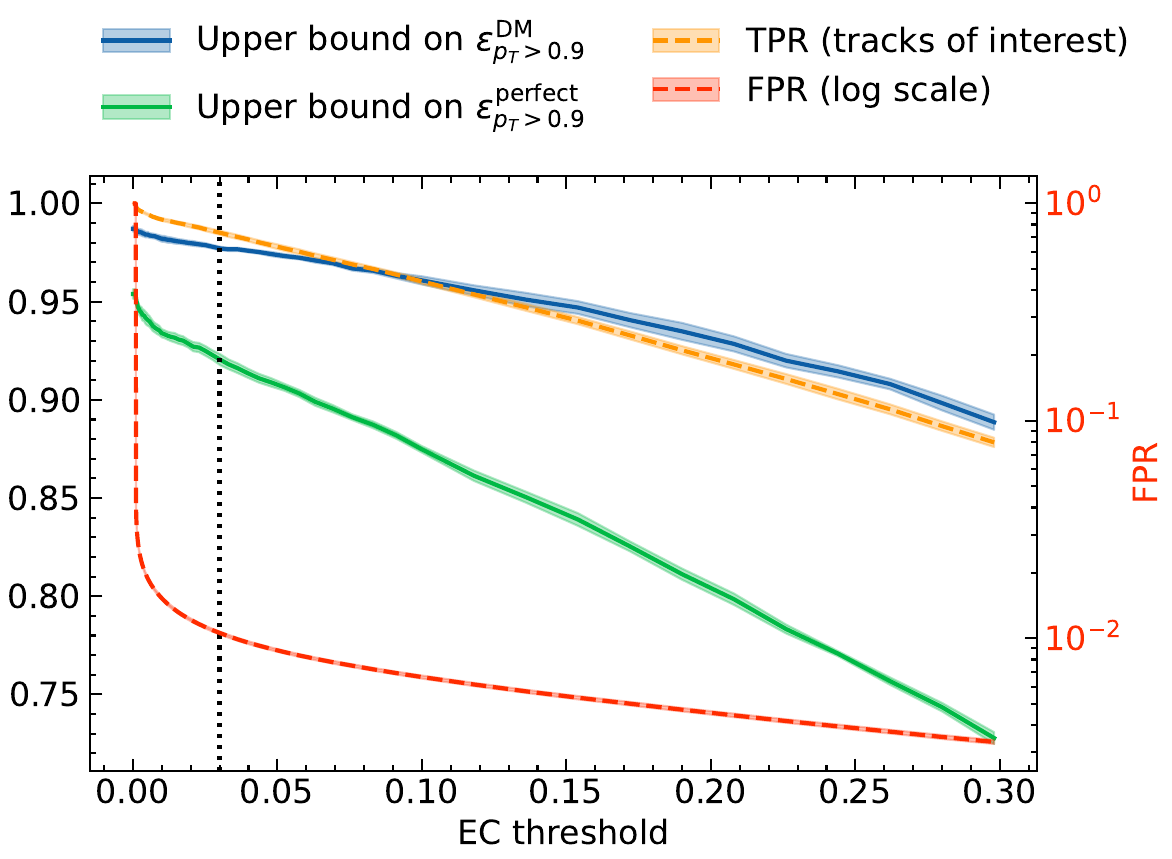}
    \caption{Upper efficiency bounds}
    \label{fig:ef_performance:upper}
    \end{subfigure}
    \begin{subfigure}{0.47\linewidth}
    \includegraphics[height=4.75cm]{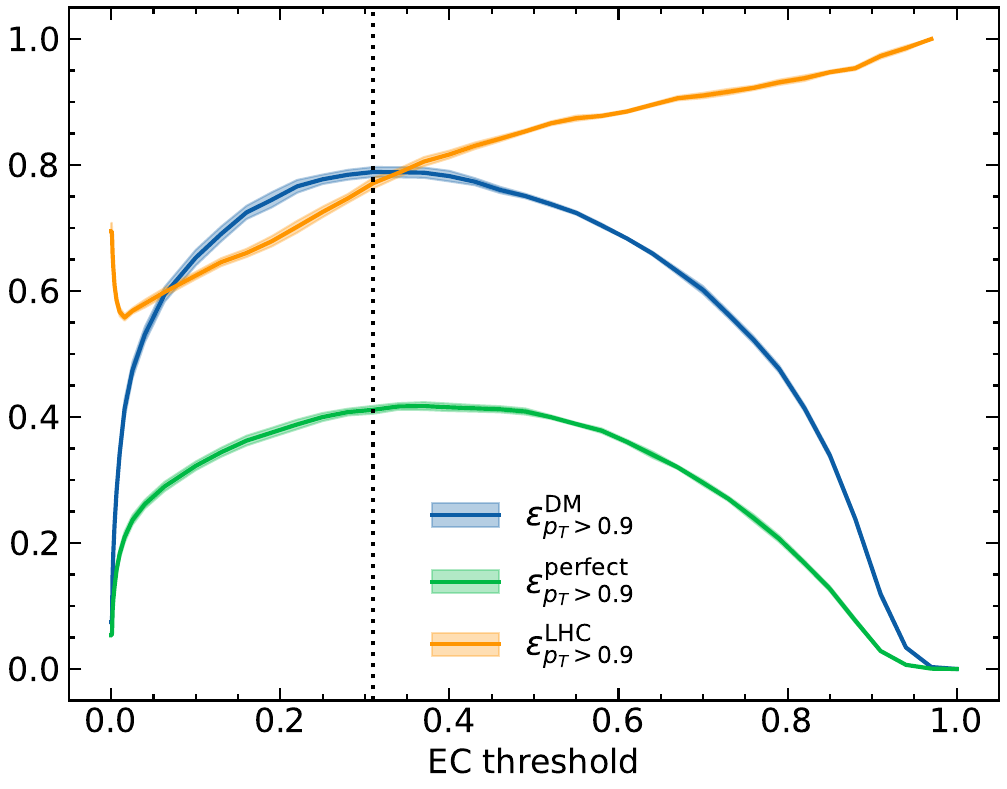}    
    \caption{Lower efficiency bounds}
    \label{fig:ef_performance:lower}
    \end{subfigure}
    \vspace{-1cm}
    \caption{Performance of the EC applied during edge construction. The left plot shows the true positive and false positive rates together with approximate upper bounds on the recoverable performance based on a perfect EC. The right plot shows the achievable tracking performance by applying a cut on the EC and identifying tracks with connected components of the resulting subgraph. Note that $\effLhcPt{0.9}$ is not strictly monotonous because it is normalized to the number of reconstructed tracks rather than particles.}
    \label{fig:ef_performance}
\end{figure}

The classifier achieves a ROC AUC of $93.3\%$ when evaluated on all tracks and $99.8\%$ when evaluated on all tracks of interest. 
Here, \emph{tracks of interest} refers to tracks with $\pt > 0.9\,\gev$ and the \emph{reconstructable} constraints described in~\autoref{sec:metrics}. 
To find an appropriate threshold, we calculate the perfect EC upper bounds to $\effDmPt{0.9}$ for the subgraphs satisfying $w_{ij}<w_\text{thld}$ for all edges. 
Based on~\autoref{fig:ef_performance:upper}, we set $w_\text{thld}=0.03$, resulting in $\mathrm{TPR}=48.3\%$, $\mathrm{TPR}\text{ (tracks of interest)}=98.5\%$, $\mathrm{FPR}=1.1\%$.
The perfect EC upper bounds for this threshold are $\effDmPt{0.9} \ecless 97.7\%$, $\effPerfectPt{0.9}\ecless 92.1\%$.
The purity of the graphs is $68\%$ at $89\times 10^3$ edges per graph.

We can also establish a \enquote{lower bound} on the performance of the pipeline by reconstructing tracks directly based on this stage. 
For this, we identify tracks with connected components of the aforementioned subgraphs (though with a stricter value of $w_\text{thld}$) and calculate the efficiencies.
A scan over $w_\text{thld}$ is shown in~\autoref{fig:ef_performance:lower} and shows a maximum of $\effDmPt{0.9}$ at $w_\text{thld}=0.31$ with
$\effDmPt{0.9} = 78.9\%$, $\effLhcPt{0.9} = 77.1\%$, and $\effPerfectPt{0.9} = 41.1\%$.
However, it should be noted that more elaborate probabilistic schemes to build tracks based on edge scores might surpass these numbers. 
\section{Object condensation}
Our architecture extends traditional EC pipelines with an additional step based on object condensation (OC)~\citep{Kieseler_condensation_2020}, a set of truth definitions and loss functions designed to cluster hits belonging to the same object and regress the properties of the reconstructed objects. OC has been extensively validated in applications to calorimetry ~\citep{Kieseler_condensation_2020,qasim_multi-particle_2021,qasim2022endtoend}, but its applications to tracking have to-date been relatively unexplored. 
\subsection{Loss functions}
%
For each hit~$i$, the OC network predicts condensation strengths
$\beta_i\in\mathbb{R}$ and clustering coordinates $c_i\in\mathbb{R}^{d_c}$. 
During training, the highest-$\beta_i$ hit in each track is dubbed the track's condensation point (CP); the goal of OC is to cluster hits around their track's CP in the learned clustering coordinate space. 
The condensation strength of a track $t$ is that of its CP, i.e., $\beta^{(t)} = \max_{\{i|l_i=t\}}\ \beta_i$. The condensation strength predicted for each hit is used to calculate an un-physical \enquote{charge} defined by $q_i=\text{arctanh}^2\beta_i+q_\mathrm{min}$ (here, $q_\mathrm{min}$ is treated as a hyperparameter). The charge corresponding to a track's CP is denoted $q^{(t)}$, located at the position $c^{(t)}$. During training, the CPs for each track are used to define attractive and repulsive losses designed to produce well-separated clusters of hits belonging to the same track in the $c_i$ coordinates:
\begin{align}
    L_\mathrm{V}
    &\coloneqq \frac{1}{N}\sum_{i=1}^{N} q_i \sum_{t=1}^{N_t}\bigg(\delta_{(l_i=t)}V^\mathrm{att}_t(c_i) + s_{\text{rep}}\big(1-\delta_{(l_i=t)}\big)V^\mathrm{rep}_t(c_i) \bigg).
\end{align}
Here, $\delta_{(l_i=t)}$ is $1$ when the node's track label is $t$ and $0$ otherwise, and $s_\text{rep}$ is a hyperparameter.
The potential functions are a quadratic attractive loss and a repulsive hinge loss:
\begin{align}
    V^\mathrm{att}_t(c) \coloneqq \delta_{(\pt > 0.9)}\, q^{(t)}\,\|c^{(t)} - c\|^2, \quad
    V^\mathrm{rep}_t(c) \coloneqq q^{(t)}\max(0, 1-\|c^{(t)}-c\|),
\end{align}
where $\delta_{(\pt > 0.9)}$ (defined as in~\autoref{eq:ef_loss}) excludes hits from noise or low-$\pt$ particles from the attraction.
An additional loss term $L_\beta$ is designed to encourage a unique CP for each track and suppress the condensation strengths of noise hits: 
\begin{equation}
    L_\beta \coloneqq \frac{1}{N_t} \sum_{t=1}^{N_t}(1-\beta^{(t)}) + s_\text{B}
    \frac{\sum_{i=1}^N \beta_i\, \delta_{(l_i=0)}}{\sum_{i=1}^{N} \delta_{(l_i=0)} }.
\end{equation}
Here, $s_B$ is a hyperparameter that controls the strength of noise suppression.
All loss terms are finally combined as $L\coloneqq L_V + s_\beta L_\beta$. 
For the results in this paper, we choose $s_\text{rep}=0.6$, $s_\beta=0.004$, $q_\text{min}=0.34$, and $s_\text{B}=0.09$.
To reduce the memory footprint of the loss functions, the graphs are split in 32 sectors during training.
\subsection{Model}
The GNN that is doing the heavy lifting of this tracking pipeline is built from \emph{interaction network} layers~\cite{battaglia_interaction_2016} with residual connections in the node updates.
Node and edge features are first encoded, $x_i^{(1)} \coloneqq W^\text{enc}_\text{node}x_i$, $e_{ij}^{(1)} \coloneqq W^\text{enc}_\text{edge}e_{ij}$ , where $(i, j)\in I$, $W^\text{enc}_\text{node}\in\mathbb R^{192\times 14}$, $W^\text{enc}_\text{edge}\in\mathbb R^{192\times 4}$. 
Then, five iterations of message passing are performed with 
\begin{align}
\begin{split}
    e_{ij}^{(k+1)} &\coloneqq \bigl(\Phi^{(k+1)}\circ \relu\bigr)\Big(\bigl[x_i^{(k)}, x_j^{(k)}, e_{ij}^{(k)}\bigr]\Big),\\
    x_i^{(k+1)} &\coloneqq \sqrt\beta\, \Psi^{(k+1)}\Bigl(\bigl[\relu\bigl(x_{i}^{(k)}\bigr), {\textstyle\sum\nolimits_{j\in\mathcal N_{i}}} e_{ij}^{(k+1)}\bigr]\Bigr) + \sqrt{1-\beta}\, x_i^{(k)}.
    \label{eq:resin}
\end{split}
\end{align}
Here, $\Phi$ and $\Psi$ are FCNNs with $\relu$ activations, a layer width of 192, and one hidden layer.
$\beta$ has been chosen to be $0.2$.
Finally, the outputs are decoded as $c_i \coloneqq W^\text{dec}_{\text{c}} \relu(x^{(6)})$ (clustering coordinates), $\beta_i \coloneqq \sigma\bigl(W^\text{dec}_\beta \relu(x^{(6)})\bigr)$ (condensation likelihoods), where $\sigma$ is the logistic function, and $W_\text{c}^\text{dec}\in \mathbb R^{192\times 24}$, $W^\text{dec}_\beta\in \mathbb R^{192\times 1}$.  
The total number of parameters of this model is $1.9\times 10^6$. 
\subsection{Postprocessing and results}
\begin{figure}
    \centering
    \begin{minipage}[t]{0.48\textwidth}
    \centering
    \includegraphics[height=4.7cm]{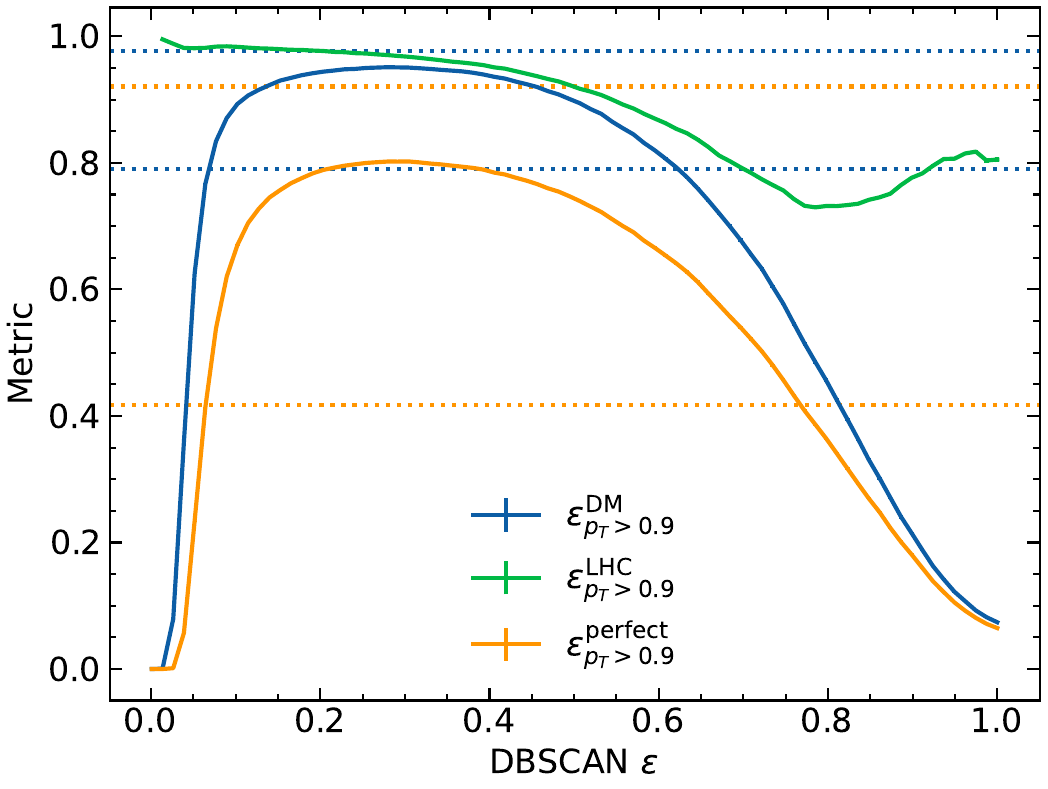}
    \caption{Optimizing the DBSCAN hyperparameters to obtain the maximum performance. The dashed lines are the upper and lower bounds after the EF step for reference (see~\autoref{fig:ef_performance}).}
    \label{fig:oc_performance_vs_eps}
    \end{minipage}
    \hspace{0.02\textwidth}
    \begin{minipage}[t]{0.48\textwidth}
    \includegraphics[height=4.7cm]{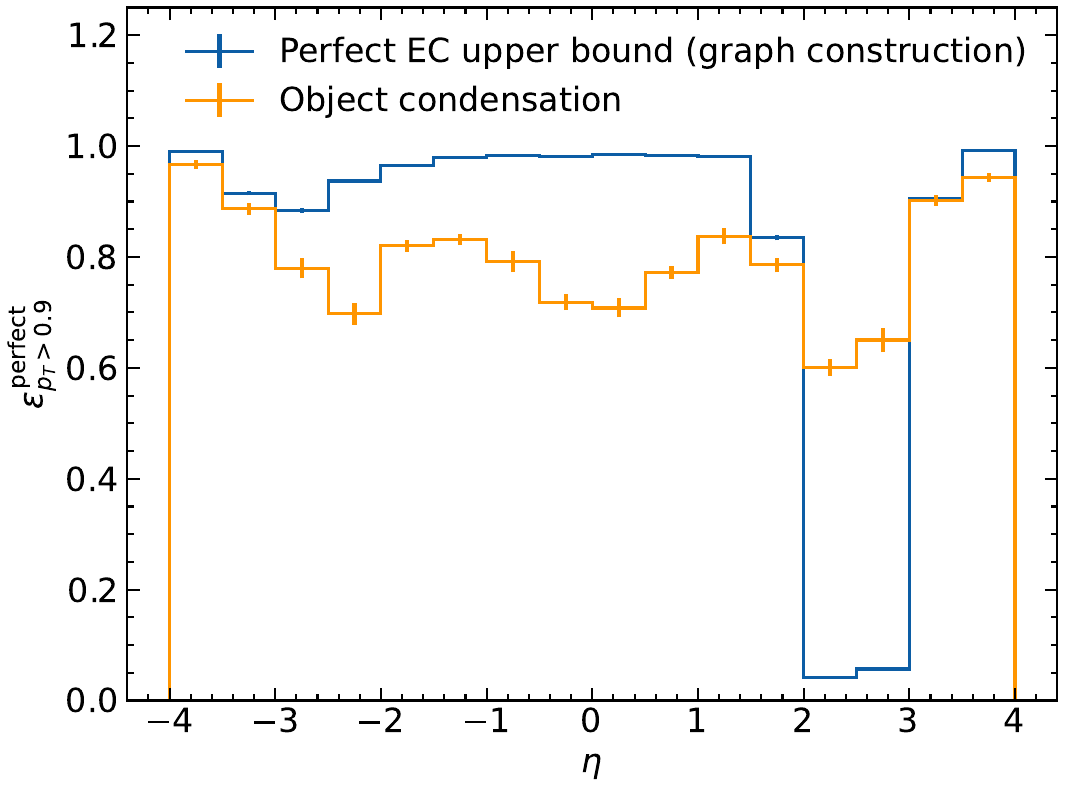}
    \caption{An OC pipeline outperforms a perfect EC when edges between the barrel and the right endcap have been removed.}
    \label{fig:oc_with_broken_gc}
    \end{minipage}
\end{figure}

\begin{figure}
    \centering
    \includegraphics[height=4.8cm]{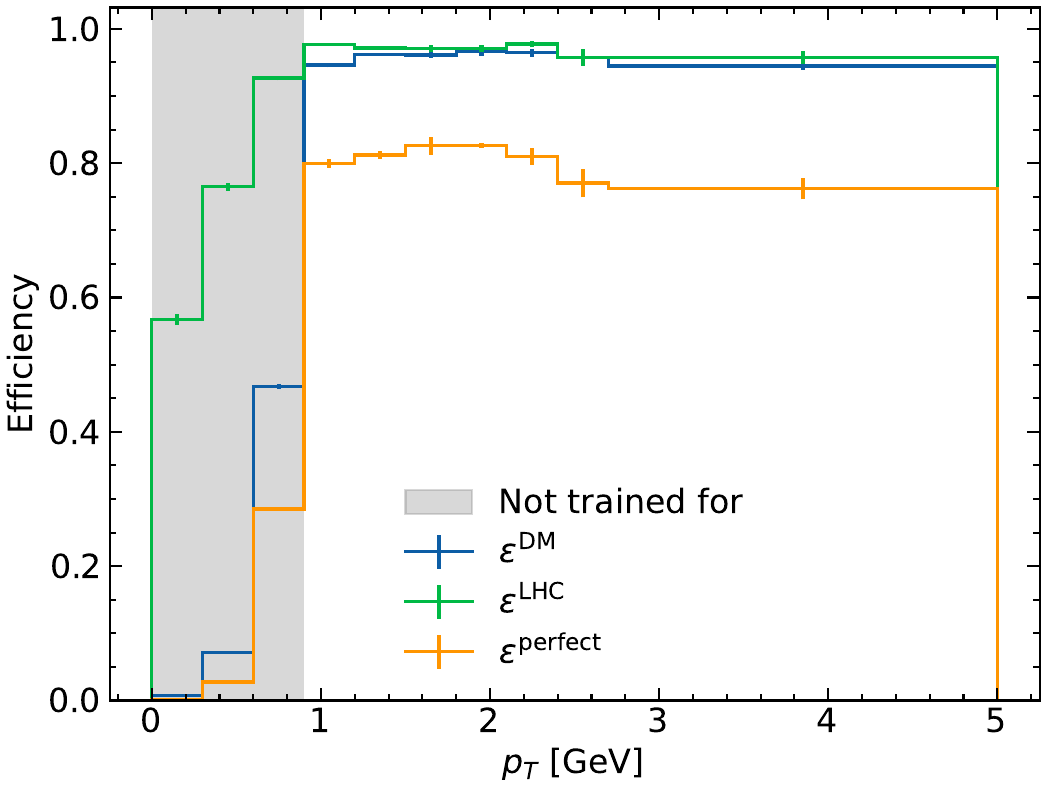}
    \includegraphics[height=4.8cm]{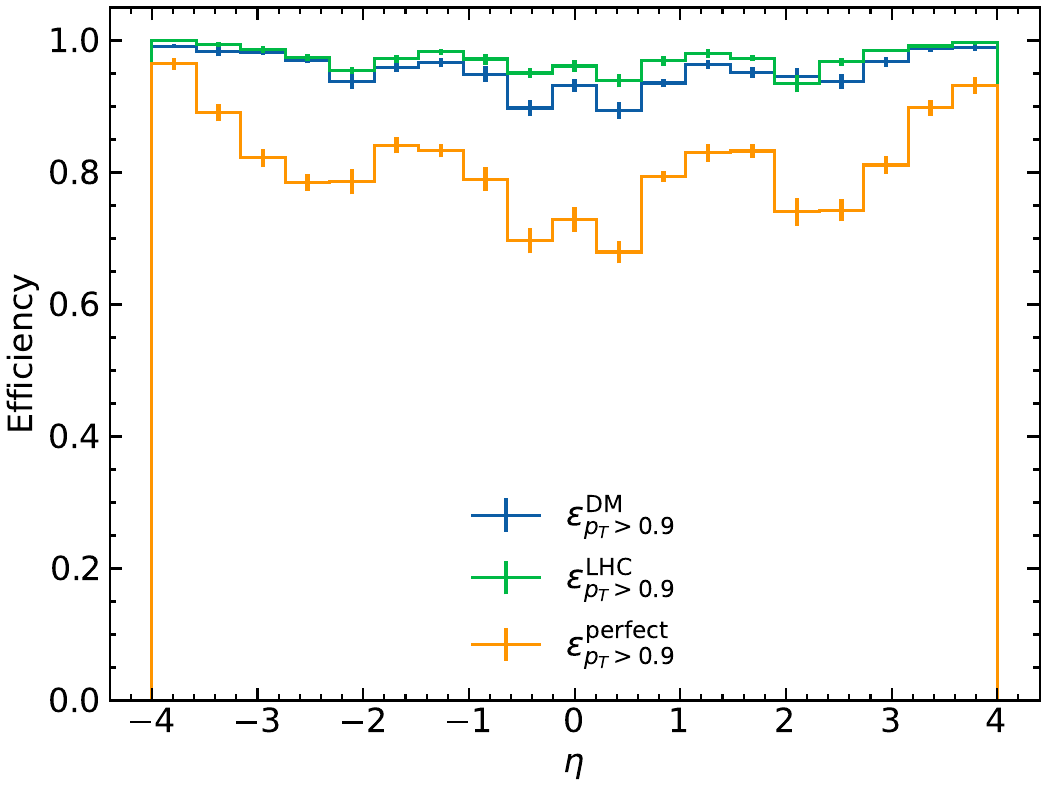}
    \includegraphics[height=4.7cm]{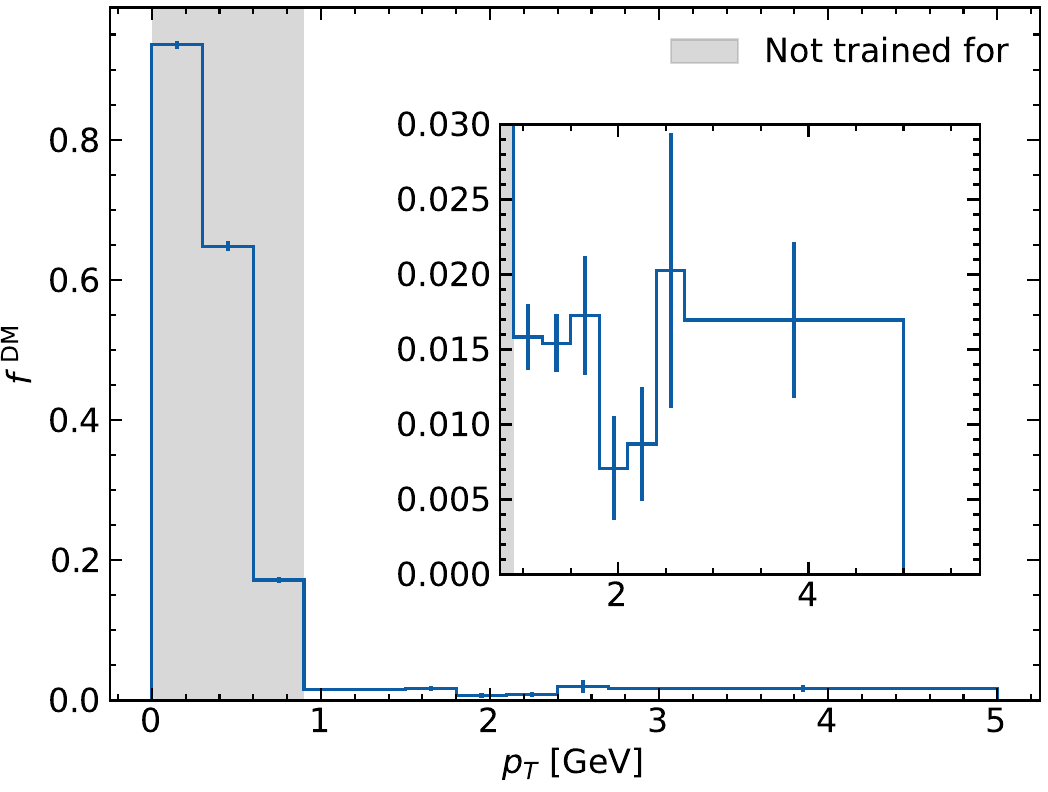}
    \includegraphics[height=4.7cm]{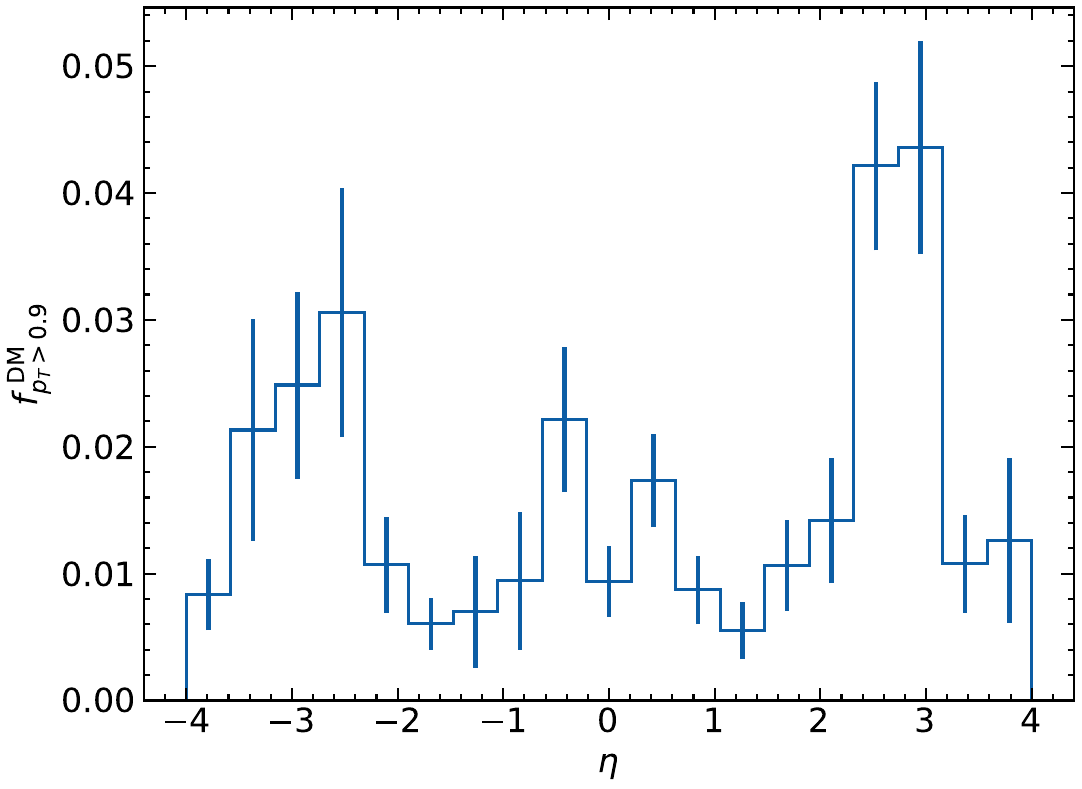}
    \caption{Tracking performance in bins of $\pt$ and $\eta$.}
    \label{fig:oc_performance_pt_eta}
\end{figure}

%
Hit clusters produced in the OC clustering space must be rendered by a downstream algorithm.
For this, we use the Density-Based Spatial Clustering of Applications with Noise (DBSCAN), an iterative clustering algorithm that has two parameters, $\epsilon$ (defining the size of the neighborhood of a point that is considered when merging clusters), and $k$ (minimum number of points within a neighborhood for the points to be considered a \emph{core point})~\cite{ester1996density}. 
For this application, $k=1$ is optimal; maximizing $\effDmPt{0.9}$ vs $\epsilon$ yields $\epsilon=0.279$ (see~\autoref{fig:oc_performance_vs_eps}).
With this, we obtain $\effDmPt{0.9}=95\%$, $\effLhcPt{0.9}=97\%$, $\effPerfectPt{0.9}=80\%$ and $f_{\pt>0.9}=1.7\%$. 
All metrics are presented vs $\pt$ and vs $\eta$ in~\autoref{fig:oc_performance_pt_eta}.

In a side study, we have also tested the ability of the OC network to reconstruct tracks with missing edges after graph construction or edge filtering. 
For this, all edges between the barrel and the right endcap have been removed after graph construction, limiting the upper bound for $\effPerfectPt{0.9}$ for a perfect EC to almost zero for tracks with $2<\eta<3$.
However, as OC is using edges only as a means to exchange information, it is not subject to this upper bound.
Indeed, the OC pipeline achieves $\effPerfectPt{0.9}=60\%$ in this region.
This is shown in~\autoref{fig:oc_with_broken_gc}.
\begin{figure}
    \centering
    \includegraphics[height=4.7cm]{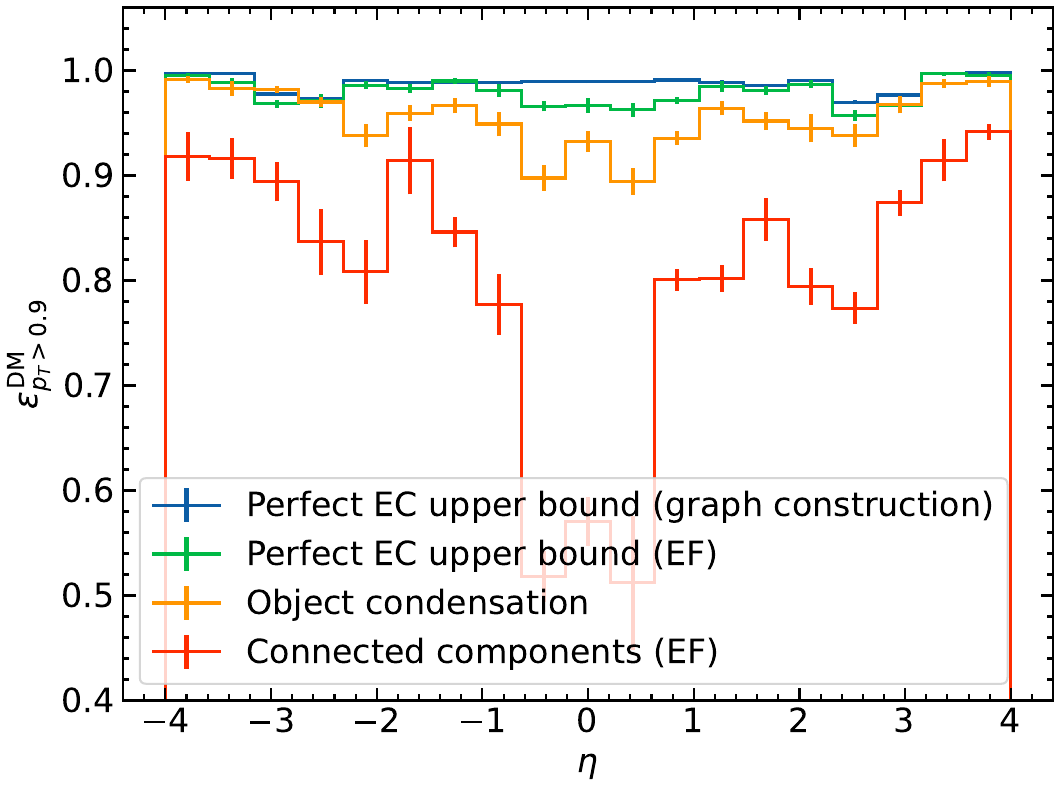}
    \includegraphics[height=4.7cm]{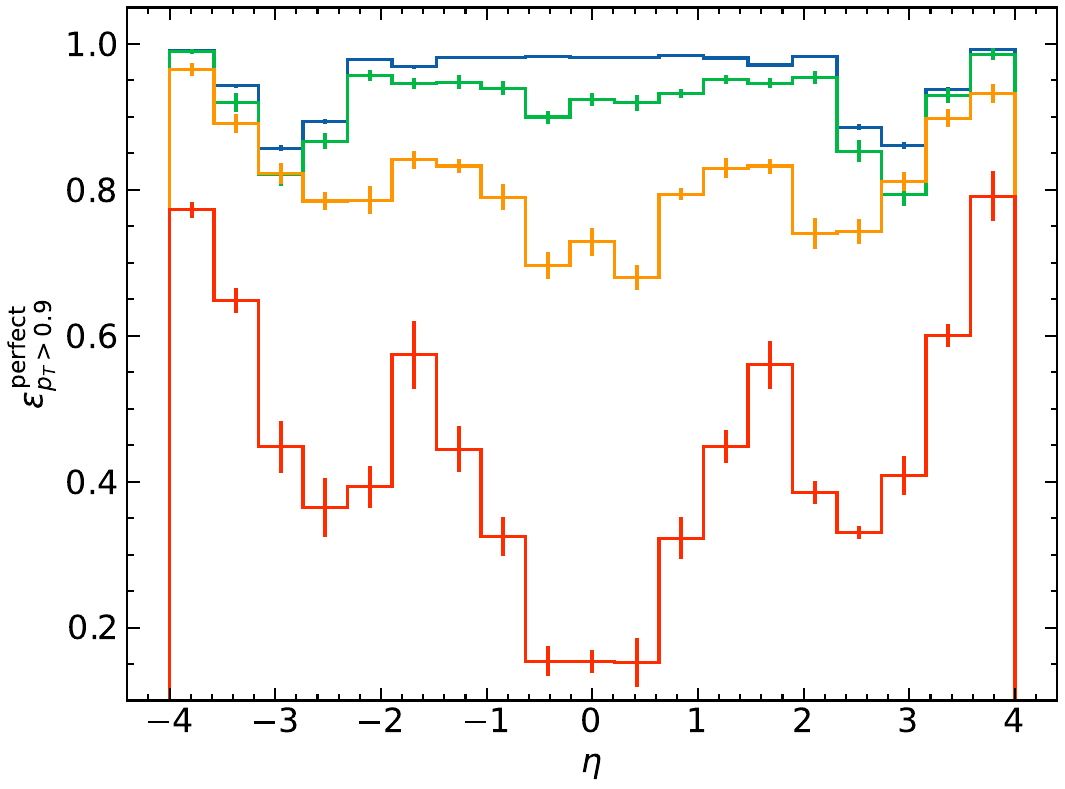}
    \caption{Comparing the performance of the OC pipeline with the upper and lower bounds introduced in this paper.}
\end{figure}

\section{Summary}
This paper presents the first GNN-based charged particle tracking pipeline that uses the OC approach to reconstruct tracks in the worst-case pileup conditions expected at the HL-LHC. 
Our pipeline shows excellent performance with respect to several metrics when applied to the pixel detector of the TrackML dataset. 
We also demonstrate that OC approach can join partial tracks that are not connected by any of the edges used for message passing, allowing it to outperform algorithms that solely rely on the output of an EC in certain scenarios.
This suggests that the use of OC at the output stage of EC-based pipelines may lead to a boost in performance. 
Future applications of OC may also allow for the regression of track parameters, for example transverse momentum, as part of an architecture capable of rendering tracks and performing preliminary fits in one shot. This incorporation of track physics may well also lead to a more robust model.

All results were produced with the open-source project \cite{Lieret_gnn_tracking_An_open-source} that implements various OC tracking architectures in a modular and extensible Python package. 
\bibliography{main}
\end{document}